\documentclass[preprint,12pt,authoryear]{elsarticle}

\usepackage{amssymb}
\usepackage{amsmath}
\usepackage{subcaption}
\usepackage{url}
\usepackage{multirow}
\usepackage{bm}

\journal{Ecol. Model}

\begin{document}

\begin{frontmatter}



\title{Household scale Wolbachia release strategies for effective dengue control} 


\author[inst1]{Abby Barlow}

\affiliation[inst1]{organization={Department of Mathematical Sciences},
            addressline={The University of Bath}, 
            city={Bath},
            postcode={BA2 7AY}, 
            state={Somerset},
            country={United Kingdom}}

\author[inst1]{Ben Adams}
\begin{abstract}
The release of Wolbachia-infected mosquitoes into Aedes aegypti infested areas is a promising strategy for localised eradication of dengue infection. Ae aegypti mosquitoes favour urban environments as breeding habitats, so are often found in and around houses. Therefore, it is likely that they will infect members of the households that they reside around. Since population groupings within households are small, stochastic effects become important. Despite this, little work has been carried out to investigate the outcome of releasing Wolbachia-infected mosquitoes at a household scale, either from an empirical and theoretical stand point. In previous work, we developed and analysed a stochastic (continuous time Markov chain) model for the invasion of Wolbachia-infected mosquitoes into a single household containing a population of wildtype mosquitoes. In the present study, we extend our framework to a connected community of households coupled by the movement of mosquitoes. We use numerical results obtained via Gillespie's stochastic simulation algorithm to investigate optimal strategies for the release of Wolbachia-infected mosquitoes carried out at either the community or the household scale. We find that household scale releases can facilitate rapid and successful invasion of the Wolbachia-infected mosquitoes into the household population and then into the wider community. We further explore the impact of regular household scale releases of Wolbachia-infected mosquitoes for a range of compositions for the release population, time intervals between releases and proportion of households participating in the releases. We find that a single release household can provide sufficient protection to the entire community of households if releases are carried out frequently for a number of years and a sufficient number of females are released on each occasion.
\end{abstract}

\begin{highlights}
\item We developed a simulation model for the movement of mosquitoes between households.
\item We investigated the release of Wolbachia-infected mosquitoes at both the household and community scale.
\item Household scale releases result in rapid and successful invasion into the household and the wider community.
\item When regular releases are carried out in a single household it can provide protection for the entire community of households if releases are carried out frequently for a number of years and a sufficient number of females are released on each release event.
\end{highlights}


\end{frontmatter}


\section{Introduction}\label{sec:intro}
Dengue is a common vector-borne disease which is transmitted to humans via the bite of an infected mosquito \cite{WHO}. It is a neglected tropical disease that is present in over $100$ countries and accounts for around $40,000$ deaths worldwide each year \citep{hasan2016dengue,bhatt2013global,zeng2021global}. The case burden is estimated to range between $100$ and $400$ million per year \citep{zeng2021global}. The development of vaccinations against dengue has been challenging. Dengvaxia is currently the only fully licensed vaccination and is only recommended for individuals who have already been infected with at least one serotype \citep{WHOvac}. 
Therefore, efforts to prevent dengue infection often target the vector populations. The most common carrier, the Aedes aegypti mosquito, inhabits urban and suburban areas, breeding in artificial water containers. These breeding sites are often found in and around homes. Insecticides are sprayed both indoors and outdoors to control the population. However, the method is unsustainable in the long-term since continual spraying is often needed to completely exterminate the mosquito population. This can result in the mosquitoes building up resistance to insecticides. 

Another well established approach to mosquito control is to introduce a large number of sterilised male mosquitoes into the local population. Often this is implemented by passing small levels of ionizing radiation over the population to be introduced \citep{strugarek2019use}. A more recent sterile insect approach is based on the release of genetically modified male mosquitoes who carry a lethal dominant gene; this technique is known as RIDL \citep{dorigatti2018using}. An alternative control method utilises the intracellular bacteria, Wolbachia. Wolbachia does not cause any harm to humans and is found naturally in around $60\%$ of insects \citep{cdc}. Infection alters the biology of Ae aegypti in numerous ways that can reduce the prevalence of infection. Firstly, with certain Wolbachia strains developed for Ae aegypti transinfection, a Wolbachia-infected male and a wildtype female cannot produce viable offspring. In contrast, the offspring of an infected male and an infected female are viable and, with a certain probability, Wolbachia positive. This mechanism is called cytoplasmic incompatibility (CI) \citep{dorigatti2018using} and affords infected females a reproductive advantage over wildtypes. Secondly, Wolbachia infection can impose fitness costs on the mosquitoes, such as decreased lifespan and fecundity and prolonged larval development. Finally, Wolbachia infection has been shown to reduce the ability of mosquitoes to transmit infection via their salivary glands \citep{hughes2013modelling}. 

Many empirical studies have investigated how Wolbachia-infected mosquitoes invade wildtype populations and can be used as a control measure against dengue \citep{ross2021designing,dos2022estimating,o2018scaled,hoffmann2014invasion,dufault2022disruption}. Cage experiments have shown that certain strains are able to invade wildtype populations and establish in large numbers \citep{mcmeniman2009stable,walker2011w}. Numerous field releases have also been implemented. In a small community near Cairns, Australia, in $2012$ almost $300,000$ mosquitoes infected with the wMel Wolbachia strain were released over $10$ weeks. The Wolbachia-infected mosquito frequency peaked at $90\%$ in the local population \citep{hoffmann2011successful}. 
Field releases of the wMelPop strain (distinct from the wMel strain) were implemented in Tri Nguyen, Vietnam (2013) \citep{nguyen2015field}. In this study, they released pupae instead of adult mosquitoes over a period of $23$ months. The Wolbachia frequency exceeded $90\%$ by the end of the release period. However, approximately $40$ to $60$ weeks later the frequency dropped to $20\%$, suggesting that the strain's high fitness cost renders it unsustainable in the absence of regular re-introduction.
More recently, large scale releases have been carried out in Indonesia, Vietnam, Australia, Columbia and Brazil. Unlike previous studies, these experiments focussed on determining the viability of the wMel strain as a dengue control measure \citep{dorigatti2018using,dufault2022disruption,o2018scaled,dos2022estimating}. They found that releasing mosquitoes infected with this Wolbachia strain successfully interrupted focal dengue virus transmission and reduced the incidence of human infection in the area.

Mathematical modelling has played an important role in the development of Wolbachia control by characterising the eco-epidemiological dynamics and untangling the relationships observed in data. Many previous models have used ODEs to describe the changes in density of the wildtype and Wolbachia-infected mosquito populations \citep{hughes2013modelling,ndii2015modelling}. These models often assume homogeneous, well-mixed populations and may account for additional aspects such as multiple mosquito life stages and seasonal fluctuations \citep{ndii2015modelling}. These relatively simple models incorporate human infection dynamics and use the infection basic reproduction number $R_0$ as an indicator for dengue control. The authors of \cite{hughes2013modelling} found that when Wolbachia infection is able to invade and persist in the population, it has the potential to eliminate dengue when $R_0$ is low. However, \citep{ndii2015modelling} analysed a model parameterised for the wMel strain and found that Wolbachia infection did not reduce mosquito lifespan sufficiently to affect population persistence.

The authors of \cite{dye2024efficacy} adapted an ODE framework to produce a metapopulation model that accounts for spatially discrete mosquito habitats coupled via mosquito movement. The population dynamics in each habitat are described by a system of ODEs and movement between habitats is modelled using dispersal kernels. They deduce that under certain conditions, the introduction of Wolbachia-infected mosquitoes into a single habitat can elicit persistent establishment across all the habitats. Other models use reaction diffusion PDEs and dispersal kernels to model the movement of mosquitoes across continuous space \citep{turelli2017deploying}. Such models focus on whether the Wolbachia-infected mosquitoes are able to diffuse over a target area. In \cite{turelli2017deploying} they found that, with the wMel strain, even very slow spatial spread can lead to area-wide population invasion within a few years of release.

Simulation based models permit a more finely detailed modelling framework. The authors of \cite{hancock2019predicting} developed a metapopulation model of mosquito habitats with heterogeneous density dependent competition effects and juvenile mosquito life stages. They found that varying the `habitat quality' spatially can significantly impact Wolbachia invasion across space. The authors of \cite{pagendam2020modelling} simulated a Markov process model accounting for multiple mosquito life stages and multiple releases. They found that under a full mosquito elimination strategy, dynamically adapting male Wolbachia-release sizes can reduce cost and the risk of releasing females.

In a previous study \cite{barlow2025analysis} we introduced a stochastic mathematical model for the dynamics of Wolbachia invasion in a single household (or a population of unconnected households) using a continuous time Markov chain framework. We
examined the relationship between the invasion probability and the Wolbachia-infection population proportion under the stochastic dynamics associated with small households, and compared this with corresponding deterministic invasion thresholds. Here we extend the model to consider a community of households via stochastic simulation methods. Each household contains (within it or close by) a breeding habitat for Ae aegypti mosquitoes. We assume mosquitoes bite the human members of their household, so that the invasion dynamics unfold stochastically in the metapopulation composed of a large number of small populations connected by the movement of mosquitoes. This framework allows for localised invasion, reversion and reinvasion, heterogeneously across the community. We use this model to investigate the factors that determine invasion success, and the impact of different strategies for the release of Wolbachia-infected mosquitoes.

When designing a Wolbachia release strategy, there are numerous aspects to consider. One key factor is the scale of the release. Previously, most empirical studies have focused on performing releases at a community wide scale \citep{dos2022estimating,hoffmann2014invasion,dufault2022disruption}. Here, one or more large populations are released in public spaces and then disperse into residential areas. An alternative is to perform releases at a household scale. Here each household is responsible for performing a small release on their own property. We have found some evidence of this strategy being carried out in practice \citep{o2018scaled}. We hypothesize that household scale releases will lead to a faster invasion of the Wolbachia-infected mosquitoes. Another key factor is the composition of the release population. Releasing female Wolbachia-infected mosquitoes can result in replacement of the wildtype population by a Wolbachia-infected population if the fitness costs of Wolbachia infection are not too high. This wildtype replacement outcome may be desirable since Wolbachia-infection reduces the transmission competence of mosquitoes for dengue and some other viruses. However, releasing female mosquitoes may be undesirable if it temporarily increases the vectorial capacity. In contrast, releasing only male Wolbachia-infected mosquitoes can result in suppression of the total mosquito population size, possibly to the point of eradication \citep{ross2021designing}. This outcome may appear more desirable than replacement but mosquito-free households are at immediate risk of reinvasion from regional wildtype populations. A third possibility is a strategy in which both male and female infected mosquitoes are released in ratios that vary over time.

Here, we consider a population of households connected by the movement of mosquitoes. Female Ae aegypti mosquitoes typically move less than $100$m in their lifetime. Therefore it may be appropriate to assume they spend all or most of their time within the same household. In contrast, males have been observed to rapidly disperse up to $400$m from their initial release site \citep{trewin2021mark},  though physical barriers such as residential blocks are thought to impede their movement \citep{trewin2021mark} and field studies have estimated the mean distance travelled over their lifetimes to range between $196$m and $294$m \citep{juarez2020dispersal,marcantonio2019quantifying,trewin2020urban}. These contrasting behaviours are likely in part because females seek regular blood meals in addition to places to rest for oviposition, whereas males spend most of their time searching for a mate. 

We use a stochastic simulation framework \citep{gillespie1977exact} to explore the efficacy of a range of single and regular release scenarios. We focus on several key aspects of release design: (1) the scale the release is performed at; (2) the number and sex ratio of the released Wolbachia-infected mosquitoes; (3) the time between releases and (4) the proportion of households participating in releases. In Section \ref{sec:model}, we define our model framework, and the release scenarios we carry out. In Section \ref{sec:results}, we present our simulation results. In Subsection \ref{subsec:com-vs-hhold} we study the efficacy of performing a single community wide release in contrast to highly localised releases in each household. In Subsection \ref{subsec:male} we consider household and community scale releases composed only of male Wolbachia-infected mosquitoes. In Subsection \ref{subsec:sex-ratios}, we analyse the impact of the number and ratio of male and female Wolbachia-infected mosquitoes released into half of the households in the community in a single release event. In Subsection \ref{subsec:single-hhold} we investigate the outcome of regularly releasing Wolbachia-infected mosquitoes into a single household.

We believe that this study will be valuable in informing effective and efficient localised Wolbachia release strategies as a control against dengue infection and other vector-borne diseases.

\section{The model} \label{sec:model}
\begin{figure}
    \centering
    \includegraphics[width=0.5\linewidth]{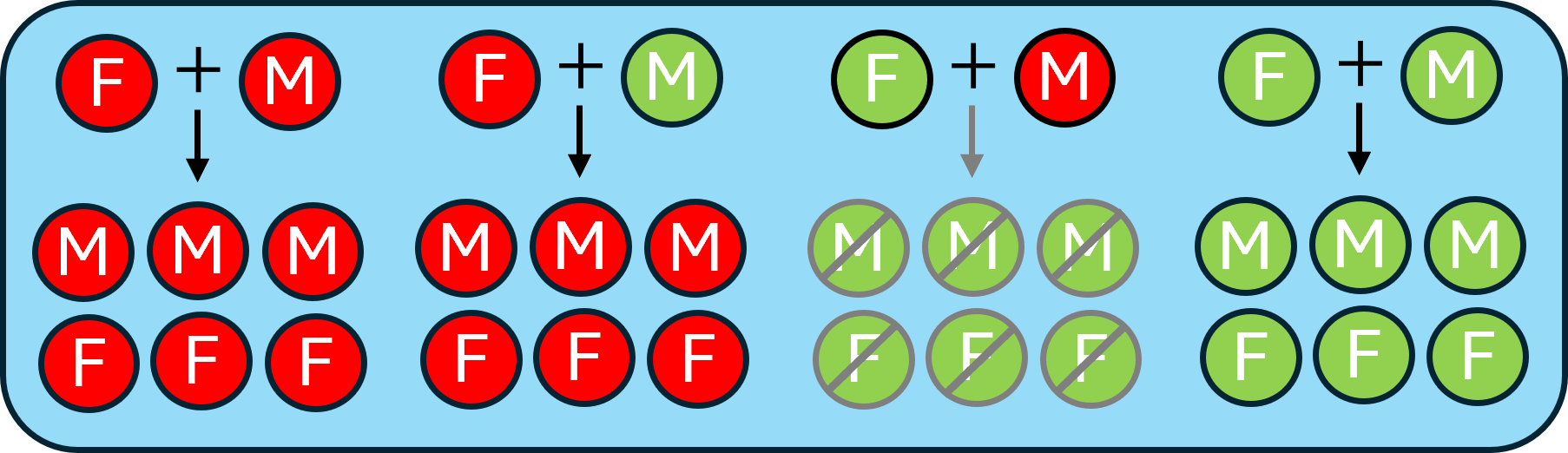}
    \caption{Schematic of the CI effect depicting the possible reproductive outcomes in the mating interactions between wildtype and Wolbachia-infected male and female mosquitoes. The top row of circles represent the male (M) and female (F) mosquito in each possible mating interaction. The circles below the arrows represent the offspring produced. Red circles correspond to Wolbachia-infected mosquitoes and green circles correspond to wildtype mosquitoes. The crossed grey outlined circles represent unviable offspring.}
    \label{fig:CI}
\end{figure}

In this section, we describe our model framework. Let the number of male wildtype and Wolbachia-infected mosquitoes in a household be denoted by $m_M$ and $w_M$ respectively. Analogously, the numbers of female wildtype and Wolbachia-infected mosquitoes in a household are denoted $m_F$ and $w_F$. The state of each household is denoted by the resident mosquito population array $(m_M,m_F,w_M, w_F)$. A household can transition to a new state via a birth, death or movement event. The rates at which the population of male wildtype and Wolbachia-infected mosquitoes die are respectively $dm_M$ and $d'w_M=d\delta w_M$. The female rates are defined analogously. Male and female birth rates are set equal. In both cases, wildtype births occur in a given household at rate
\begin{align} \label{eq:wild_birth_rate}
    bZ_mF(Z_m + Z_w),
\end{align}
where
\begin{align}\label{eq:wild_birth_rate_part2}
    Z_m=\frac{m_M}{m_M+w_M}\left(m_F + w_F(1-v)\phi\right) + \frac{w_M}{m_M+w_M}\left(m_F(1-u) + w_F(1-v)\phi\right).
\end{align}
Wolbachia-infected births occur in a given household at rate
\begin{align} \label{eq:wolb_birth_rate}
     b Z_wF(Z_m + Z_w),
\end{align}
where
\begin{align}
    Z_w=v\phi w_F.
\end{align}
These birth rates follow a similar formulation to \cite{barlow2025analysis}, except here we account for male and female mosquitoes explicitly. In \eqref{eq:wild_birth_rate} and \eqref{eq:wolb_birth_rate}, $b$ is the maximum per capita (female) birth rate and $F(Z)$ accounts for larval density dependent mortality. $Z_m$ in the wildtype birth rate \eqref{eq:wild_birth_rate} is comprised of two parts, as detailed in \eqref{eq:wild_birth_rate_part2}. The first term is the probability that a wildtype male mates with a wildtype female, or mates with an infected female but vertical transmission does not occur. The second term is the probability that an infected male mates with a wildtype female and the offspring is viable, or mates with an infected female and vertical transmission does not occur. The offspring of a Wolbachia-infected male and a wildtype female are viable with probability $u$. The fitness cost of Wolbachia-infection on the birth rate is $1-\phi$.
For the Wolbachia-infected birth rate \eqref{eq:wolb_birth_rate}, all mating interactions with Wolbachia-infected female mosquitoes produce infected offspring if vertical transmission is successful. The schematic in Figure \ref{fig:CI} details all the different mating interactions and outcomes that can occur between the male and female wildtype and Wolbachia-infected mosquitoes via the CI effect. All parameter values used throughout this study are detailed in Table \ref{tab:params}.

\begin{table}[htbp]
\centering
\footnotesize
\renewcommand{\arraystretch}{1.3}
\caption{Parameter definitions and values. Values except $K$, $b$, $\tau$, $\rho$, $\gamma$ and $\alpha$ are from \cite{hughes2013modelling} and re-scaled to mosquitoes per $100\mathrm{m^2}$, the approximate size of a household dwelling. $K$ is based on \cite{madewell2019associations} and \cite{barlow2025analysis}; $b$ is calculated in \cite{barlow2025analysis}. The dispersal parameters $\tau$, $\rho$, $\gamma$ and $\alpha$ are estimated in this study.}
\label{tab:params}
\begin{tabular}{|p{1.5cm}|p{8.5cm}|p{4cm}|}
\hline
Parameter& Definition& Value\\
\hline
$b$& Per capita (female) birth rate for wildtype mosquitoes& $4.52d=0.54\;\;\text{day}^{-1}$ \\ [10pt]
$d$& Per capita death rate of wildtype mosquitoes& $0.12\;\;\text{day}^{-1}$\\ [10pt]
$d'=\delta d$& \multirow{2}{250pt}{Per capita death rate of Wolbachia-infected mosquitoes}& $0.12\;\;\text{day}^{-1}$\\[15pt]
 $b'=\phi b$& \multirow{2}{250pt}{Per capita (female) birth rate for Wolbachia-infected mosquitoes}&$0.52\phi=0.46\;\;\text{day}^{-1}$\\[15pt]
$u$& \multirow{2}{250pt}{Cytoplasmic incompatibility. Probability of a Wolbachia-infected male and uninfected female producing inviable offspring}& $1$\\[25pt]
 $v$& \multirow{2}{250pt}{Vertical transmission. Fraction of offspring of a Wolbachia-infected female that are infected}&$1$\\[15pt]
 $\phi$& \multirow{2}{250pt}{Reproductive cost of infection. See $b^\prime$ above}& $0.85$\\[15pt]
 $\delta$& \multirow{2}{250pt}{Mortality cost of infection. See $d^\prime$ above}& $1$\\[15pt]
 $K$ & Reproductive carrying capacity (of female mosquitoes)& $30$\\[10pt]
 $k$& Larval competition parameter&  $0.3$\\ [15pt]
 $h$& Larval competition parameter&$0.19\times 100^k=0.76$\\ [15pt]
 $\tau$ & Male mosquito household leave rate & \multirow{2}{120pt}{$0.15$ (low), $0.33$ (medium), $1$ (high) $\;\;\text{day}^{-1}$}\\[15pt] 
 $\rho$ & Male mosquito household arrival rate & \multirow{2}{120pt}{$0.15$ (low), $0.33$ (medium), $2$ (high)$\;\;\text{day}^{-1}$}\\[15pt]
 $\gamma$ & Rate mosquitoes disperse from release site & \multirow{2}{120pt}{$0.15$ (low), $0.33$ (medium), $2$ (high)$\;\;\text{day}^{-1}$}  \\[15pt]
 $\alpha$  & Rate female mosquitoes switch between households & \multirow{2}{120pt}{$0.03$ (low), $0.1$ (medium), $0.5$ (high)$\;\;\text{day}^{-1}$}\\[15pt]
 \hline
\end{tabular}
\end{table}

\subsection{The intraspecific larval density function}
The function $F(Z_m + Z_w)$ represents the effects of intraspecific competition at the larval mosquito stage. We consider $Z_m + Z_w$ here, instead of the total adult mosquito population size $m_M+m_F+w_M+w_F$, in order to account for the unhatched eggs produced from unviable births due to CI pairings, which will have no impact on the competition among larvae for resources. $F(Z_m + Z_w)$ is a monotonic decreasing function of $Z_m + Z_w$, with $F(0) = 1$. Note that the birth rates are maximised when $F(Z_m + Z_w)=1$. We discuss and implement two choices of the larval density function in \cite{barlow2025analysis}. Here, we use
\begin{align}\label{eq:hughes_F}
    F(Z_m+Z_w)=\begin{cases}\exp(-h(Z_m+Z_w)^k) & m_F+w_F < K,\\
    0 & m_F+w_F \geq K.
    \end{cases}
\end{align}
This is the heuristic function used in \cite{hughes2013modelling,dye1984models} with a small modification. The birth rate is set to zero when the number of female mosquitoes in the household reaches or exceeds the (female) reproductive carrying capacity, $K$. This convention aligns with our previous framework \citep{barlow2025analysis} and ensures the number of possible household states remains computationally tractable.

\subsection{Accounting for movement and metapopulation effects}
We consider a community of $100$ households. In each household, mosquito population dynamics occur by a stochastic birth-death process as described above. In addition, we account for the movement of mosquitoes between households. There are reported to be distinct behavioural differences between male and female Ae aegypti mosquitoes in terms of their mobility \citep{juarez2020dispersal,marcantonio2019quantifying,trewin2020urban,trewin2021mark}. Male mosquitoes disperse large distances in the search of mates. In contrast, female Aedes Aegypti tend to stay in areas with access to regular blood meals and resting sites. In order to capture these different behaviours in our model male mosquitoes leave their household at per capita rate $\rho$ and enter a single `free-living' pool. They move from this pool back into a household (chosen at random) at rate $\tau$. In contrast, female mosquitoes switch directly between households at a low rate $\alpha$. Initially we assumed female mosquitoes did not move at all. However, we found that this approximation meant that household mosquito populations were unable to recover from extinction if their last female inhabitant died. 

The low-level female dispersal produces a rescue effect in the model. The rescue effect is a phenomenon observed in metapopulation systems where the probability of localised extinction, and ultimately metapopulation extinction, is reduced by migration. Small subpopulations are subject to local extinction due to stochastic fluctuations. A strong rescue effect is said to occur when the arrival of female migrants boosts the birth rate in the population. A weak rescue effect is said to occur when the arrival of migrants re-establishes a previously extinct subpopulation. Both scenarios are possible under the framework presented in this study. Metapopulation dynamics have been extensively documented in species residing in fragmented landscapes \citep{hanski2004ecology}. The rescue effect is greatest when subpopulation dynamics are either uncorrelated or negatively correlated. Correlation in subpopulation dynamics can lead to metapopulation-wide extinction events due to synchronisation \citep{keeling2004metapopulation}.
The metapopulation as a whole, as well as its individual component subpopulations, are often much more stable and persist for longer than small populations in solitude \citep{hanski2004ecology}.

It is possible to write down a Master Equation to describe the rate of change of the household state probability using the single household approximation \citep{holmes2022approximating}. We present this work in the Supplementary Information. However, accounting for movement between households introduces nonlinearities that render the Master Equation intractable. Hence, we are unable to utilise many of the analytic techniques introduced in previous work \citep{barlow2025analysis}. Consequently, we use numerical simulation, via the Gillespie SSA \citep{gillespie1977exact} in order to gain insight into the Wolbachia invasion dynamics in a system composed of coupled household structures.

\subsection{Release framework}\label{subsec:male-fem-single-rel}
We use our model to investigate the outcomes of a range of scenarios for the release of Wolbachia-infected mosquitoes. 

For community scale releases, we make a clear distinction between Wolbachia-infected mosquitoes dispersing from the release site and male mosquitoes moving between households and by introducing a separate mosquito population class in the model, termed the `release' population. The Wolbachia-infected mosquitoes are added instantaneously to the free-living `release' population and then disperse into households at per capita rate $\gamma$. Once they enter a household, these mosquitoes become subject to the regular male and female movement models described above.

\begin{figure}
    \centering
    \includegraphics[width=0.8\linewidth]{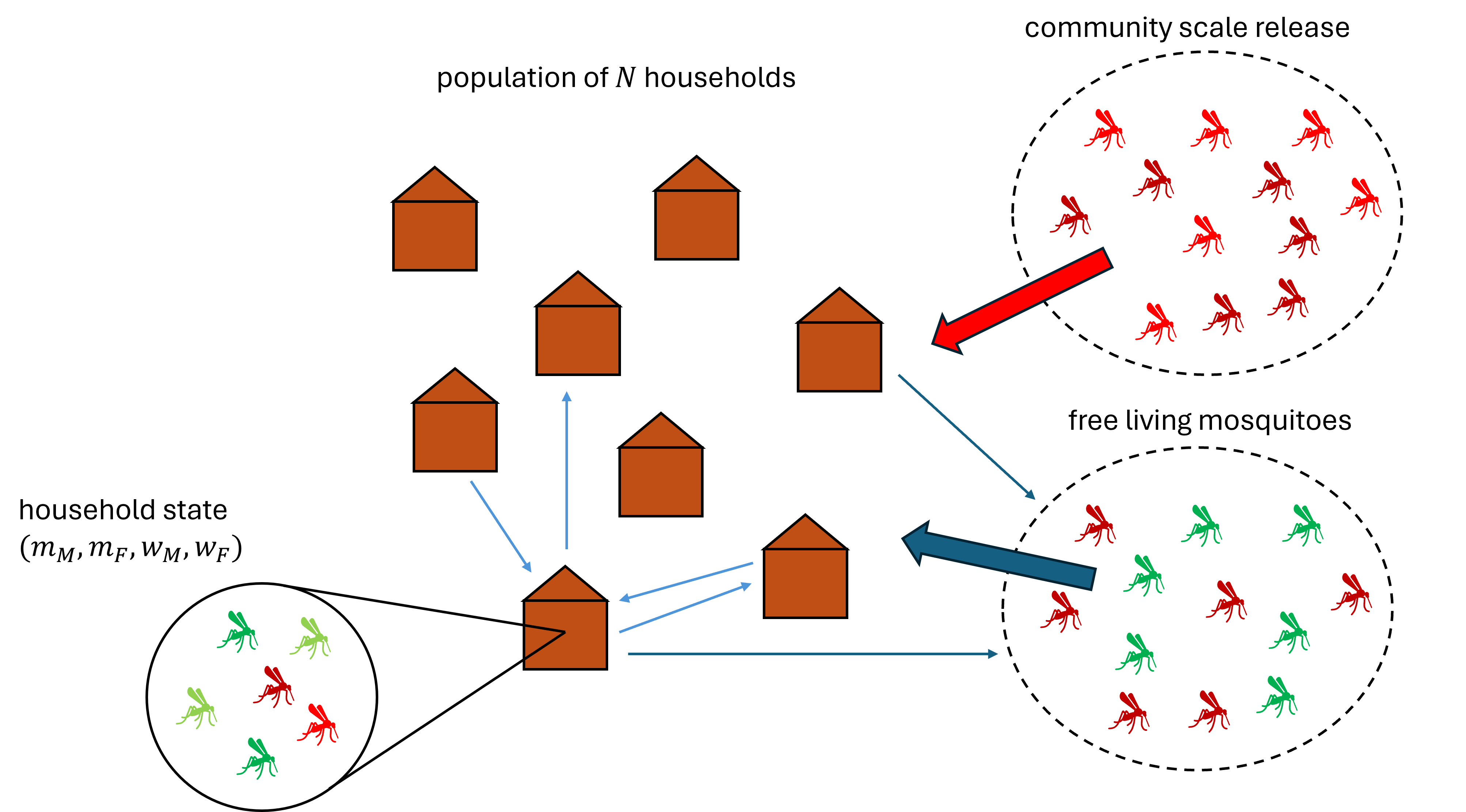}
    \caption{Schematic of the model framework. Household habitats of the mosquitoes are denoted in brown. Each household is characterised by its household state $(m_M,m_F,w_M,w_F)$, denoting the number of wildtype male $m_M$ (dark green), female $m_F$ (light green) and Wolbachia-infected male $w_M$ (dark red), female $w_F$ (light red) mosquitoes currently residing in the household. Female mosquitoes move directly between households as indicated by the lighter blue arrows. Male mosquitoes move from households into the free-living population (denoted by the lower dashed circle of mosquitoes) and then into new households, as denoted by the darker blue arrows. Wolbachia-infected mosquitoes may be released at a community scale or a household scale. The upper dashed circle of infected (red) mosquitoes represents a community scale release.}
    \label{release-schem}
\end{figure}

Figure \ref{release-schem} shows a schematic representation of the model and release strategy.  The household habitats of the mosquitoes are denoted in brown. Each household is characterised by its household state $(m_M,m_F,w_M,w_F)$, denoting the number of wildtype male $m_M$ (dark green), female $m_F$ (light green) and Wolbachia-infected male $w_M$ (dark red), female $w_F$ (light red) mosquitoes currently residing in the household. Female mosquitoes move directly between households as indicated by the lighter blue arrows. Male mosquitoes move from households into the free-living population (denoted by the lower dashed circle of mosquitoes) and then into new households as denoted by the darker blue arrows. Wolbachia-infected mosquitoes may be released at a community scale or a household scale. The upper dashed circle of infected (red) mosquitoes represents a community scale release.

\subsection{Stochastic simulation}
We solved our model by implementing Gillespie's SSA in Julia. We used a similar computational framework to \cite{barlow2025epidemiological}, employing dictionaries to keep track of the dynamic set of extant household states. We used mutable structs to define the community of households. Each household state was an instance of the struct, characterised by differing propensities of the various household events. We defined the population of free-living mosquitoes analogously using additional mutable structs. We distinguished between mosquitoes released into the free-living pool that had not yet visited a household and mosquitoes that had previously visited a household.

We considered communities composed of $100$ households. Each household was initialised with $8$ male and $8$ female wildtype mosquitoes, close to the steady state value found for a single household in \cite{barlow2025analysis}. For each realisation of the model the stochastic population dynamics were simulated for $150$ days from this starting point so that the wildtype household mosquito populations approached the quasi-stationary distribution before any Wolbachia-infected mosquitoes were released.  

The code used to produce all results in this study can be found in the GitHub repository \url{https://github.com/ahb48/Wolbachia-movement-households}. In the Supplementary Material, we reproduce analytic results from our previous work \citep{barlow2025analysis} numerically using the SSA and investigate the impact of different household population sizes and rates of dispersal between the households on the probability of a successful Wolbachia invasion into a resident wildtype mosquito population.

\subsection{Release scenarios}
In Subsection \ref{subsec:com-vs-hhold}, we investigate a single release of mixed sex Wolbachia-infected mosquitoes into a community of $N=100$ households. We consider a community scale release with $400$ male and $400$ female Wolbachia-infected mosquitoes introduced to the free-living `release' pool. We also consider a household scale release with $4$ male and $4$ female Wolbachia-infected mosquitoes introduced directly to each household in the community. We compare four release scenarios: community scale releases with low ($\gamma=0.15$), moderate ($\gamma=0.33$), and high ($\gamma=2$) release dispersal, and a household scale release also as described above. We find that household scale releases facilitate more rapid invasion of Wolbachia-infected mosquitoes, and that community scale releases with higher dispersal correspond more closely with household scale releases. In Subsection \ref{subsec:male} we investigate single male-only releases of a total of $800$ mosquitoes at the community and household scales.

In Subsection \ref{subsec:sex-ratios} we explore the impact of varying the proportion of male and female Wolbachia-infected mosquitoes in a release. We find that, due to the CI effect, this ratio affects the invasion dynamics within the release households, and the persistence of Wolbachia-infection in the metapopulation.

In practice it will rarely be possible to control the wildtype mosquito population with a single Wolbachia release event. So, in Subsection \ref{subsec:single-hhold} and the Supplementary Information we investigate the impact of releases carried out at regular intervals in a single household. We consider the protection this affords the release household, and the wider community.

\section{Results}\label{sec:results}
For each release scenario we computed $100$ independent realisations of the model. We derived several statistics from these ensembles to characterise the invasion, establishment and persistence of Wolbachia-infection at the household and community scales. 

We calculated the mean and variance of the number of mosquitoes of each type (male/female, wildtype/infected) in a household at a given time point from the set of $10,000$ households formed from all $100$ households in all $100$ realisations. The mean population size is an indicator of the extent to which Wolbachia-infected mosquitoes have replaced wildtypes in the entire metapopulation. The variance of the population size is an indicator of the uniformity of the Wolbachia invasion across the metapopulation. 

We measured the persistence time of Wolbachia infection in the entire community, or in a given household, as the proportion of the $850$ day simulation period following the release on day $150$ for which at least one Wolbachia-infected mosquito was present. For persistence within a given household, in each realisation of the model, the state of a single household in the community was monitored for the full duration of the simulation to get the duration of infection. The outcome was then averaged over $100$ realisations.

\subsection{Community versus household scale single releases} \label{subsec:com-vs-hhold}
\begin{figure}[htbp]
    \centering

    \begin{subfigure}{0.45\textwidth}
        \centering
        \includegraphics[width=\textwidth]{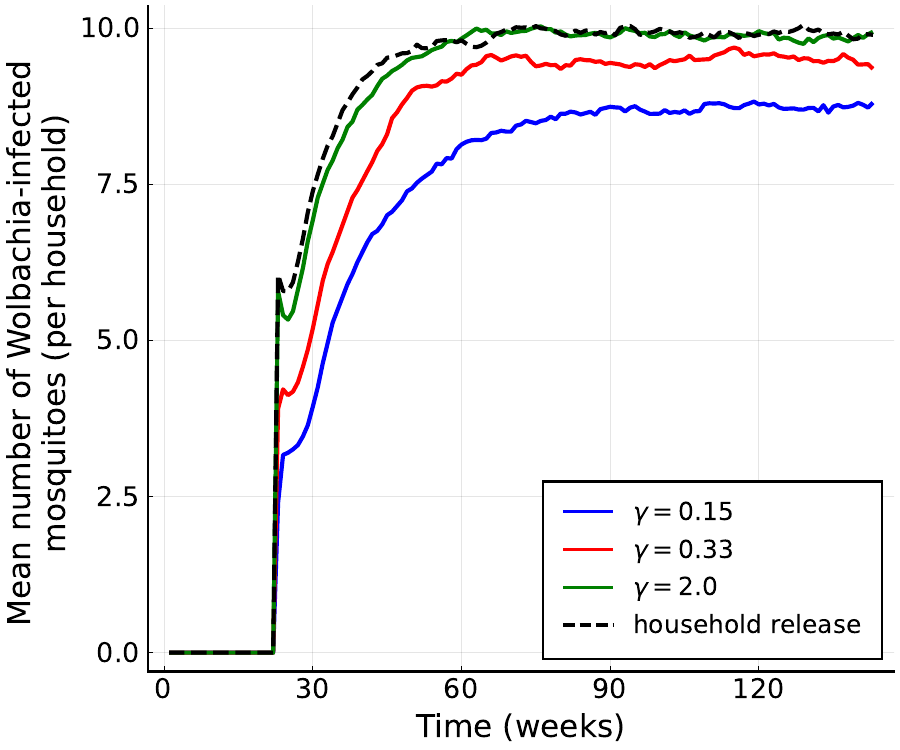}
        \caption{}
        \label{sfig3:wolb-hhold-av}
    \end{subfigure}

    \vspace{1em}

    \begin{subfigure}{0.45\textwidth}
        \centering
        \includegraphics[width=\textwidth]{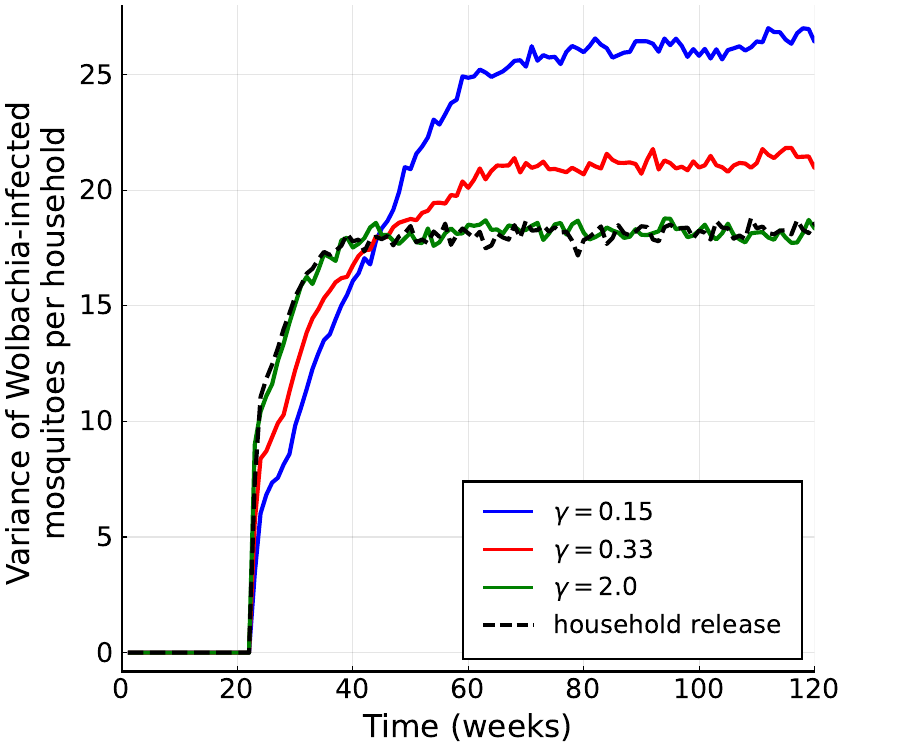}
        \caption{}
        \label{sfig1:wolb-hhold-var}
    \end{subfigure}
    \hspace{0.05\textwidth}
    \begin{subfigure}{0.45\textwidth}
        \centering
        \includegraphics[width=\textwidth]{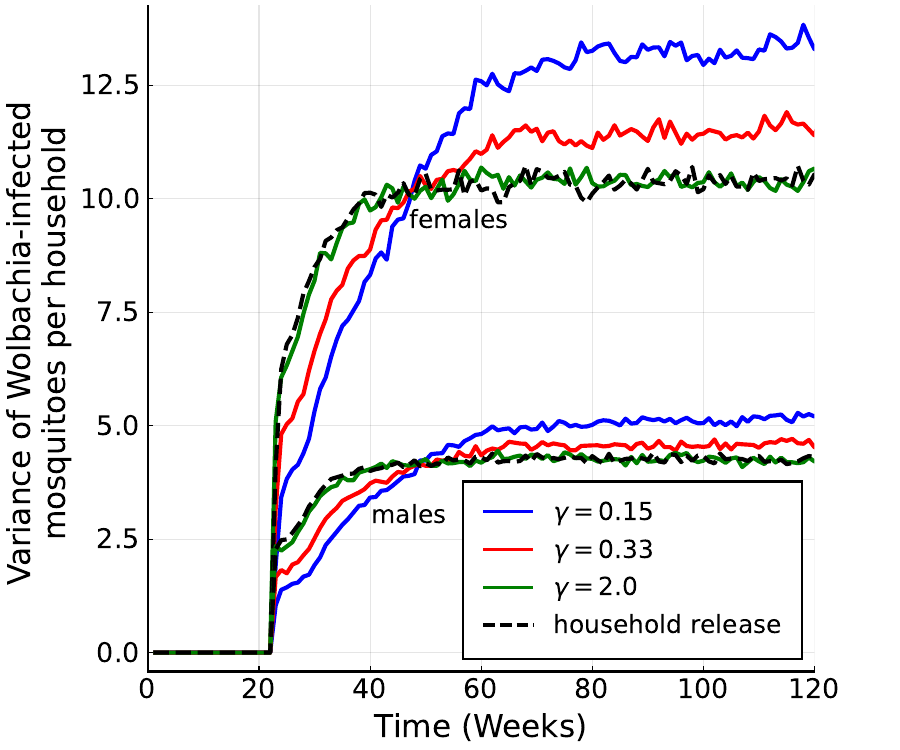}
        \caption{}
        \label{sfig2:wolb-hhol-var}
    \end{subfigure}

    \caption{Mean and variance of the number of Wolbachia-infected mosquitoes per household over time (weeks) for different release scenarios. (a) Mean for male and female mosquitoes combined. (b) Variance for male and female mosquitoes combined. (c) Variance of male and female mosquitoes separately. Statistics were calculated from $100$ model realisations each made up of $100$ households. A single release of Wolbachia-infected mosquitoes was carried out at time $150$ days. The coloured curves denote community scale releases of $800$ mosquitoes ($400$ male, $400$ female) for low (blue), moderate (red) and high (green) dispersal release rates $\gamma$. The black dashed curve denotes a household scale release of $8$ mosquitoes ($4$ male and $4$ female) in each household. All other parameter values can be found in Table \ref{tab:params}, with the intermediate values for the non-release dispersal parameters.}
    \label{fig:combined-wolb}
\end{figure}

\begin{figure}[htbp]
\centering
    \includegraphics[width=0.5\linewidth]{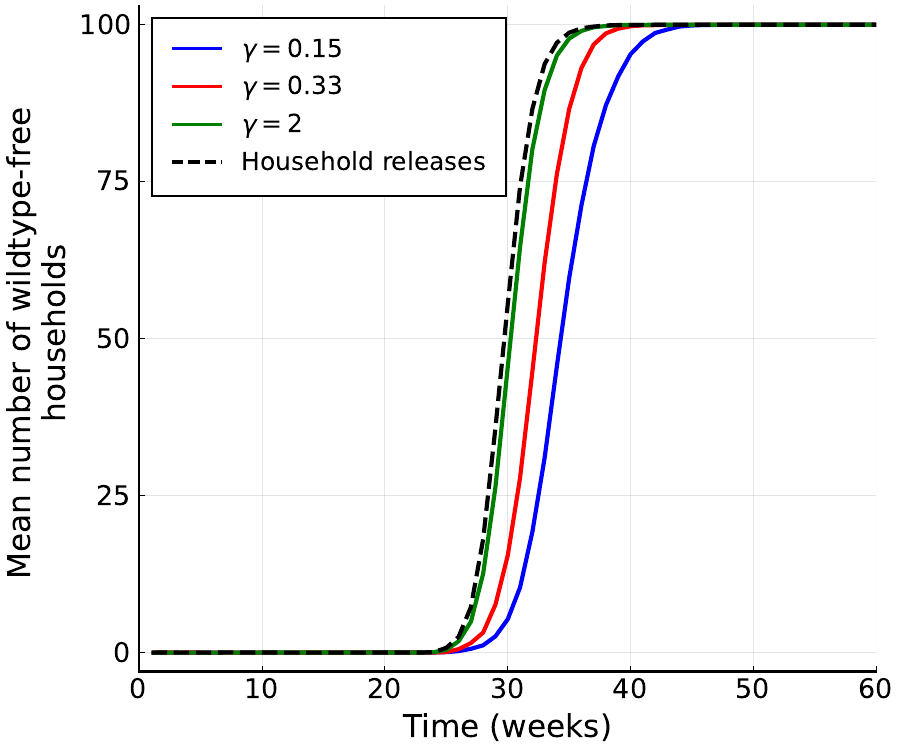}
    \caption{Mean number of households containing no wildtype mosquitoes over time (weeks) for different release scenarios. For model configuration and parameterisation see Figure \ref{fig:combined-wolb}.}
    \label{fig:no-wild-hhold-av}
\centering
\end{figure}

Figure \ref{sfig3:wolb-hhold-av} shows the mean number of Wolbachia-infected mosquitoes in a household over time for four release scenarios: household scale release and community scale release with low, moderate and high dispersal rates from the release site. Figure \ref{sfig1:wolb-hhold-var} shows the corresponding variance and Figure \ref{sfig2:wolb-hhol-var}) decomposes this variance into the male and female Wolbachia-infected subpopulations in the household. Figure \ref{fig:no-wild-hhold-av} shows how the mean number of households that contain no wildtype mosquitoes varies over time. Table \ref{tab:persistence} shows the mean persistence time of Wolbachia-infection in a household mosquito population.  

The mean population trajectories reach a quasi-steady state after around $60$ weeks. In general, lower dispersal leads to smaller Wolbachia-infected household mosquito populations with higher variance. When dispersal is low, there is a higher chance of mortality before reaching a household. Furthermore, for each household, there is less momentum to drive the initial invasion into the resident wildtype population and so more variability in the outcome. The variance in the female subpopulation is more than double that of the male subpopulation. This is because females move less frequently between households than males after the initial release dispersal. All four release scenarios lead to communities in which there are no wildtype mosquitoes in any household after $45$ weeks, although the Wolbachia-infected household populations may not yet have reached quasi-steady state. Lower dispersal is associated with slower eradication of the wildtype mosquito population and shorter mean persistence times of Wolbachia infections in household populations. 

The invasion dynamics arising from a household scale release are very similar to those of a community scale release with high dispersal ($\gamma=2$). So, in the absence of rapid dispersal of community-released mosquitoes, household scale releases may support the establishment of substantial and persistent Wolbachia-infected household mosquito populations. 

\begin{table}
    \centering
    \captionsetup{width=1\linewidth}
    \begin{tabular}{|l|l|}
    \hline
        Release scenario & Proportion of time (days)\\
        \hline
        community, $\gamma=0.15$ & $89.3\%$\\
        community, $\gamma=0.33$ & $96.4\%$\\
        community, $\gamma=2.0$ & $99.6\%$\\
        household & $99.7\%$\\
        \hline
    \end{tabular}
    \caption{Proportion of time for which there is at least one Wolbachia-infected mosquito in the household over the period between day $150$ and day $1000$. For each realisation of the model, the state of a single household was monitored for the full duration of the simulation to get the duration of infection. The results shown are mean values calculated from $100$ realisations. For model configuration and parameterisation see the caption of Figure \ref{fig:combined-wolb}.}
    \label{tab:persistence}
\end{table}

\subsection{Male only single release}\label{subsec:male}
Here we explore the same four release scenarios as described in Section \ref{subsec:male-fem-single-rel}, except each release is made up of male mosquitoes only ($800$ males released in total). 

\begin{figure}
    \centering
    \includegraphics[width=0.5\linewidth]{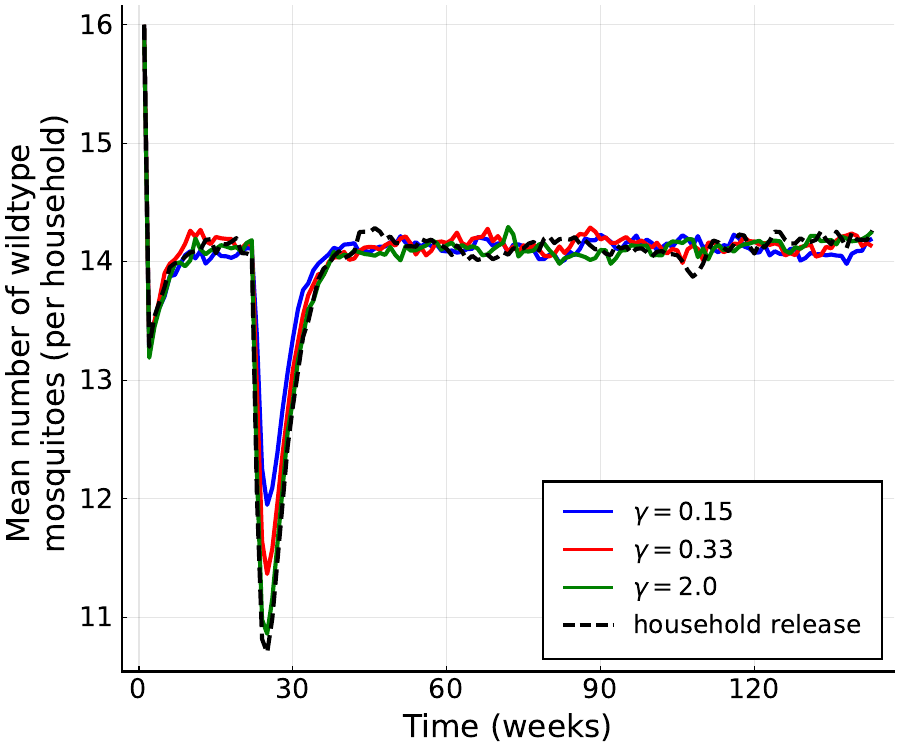}
    \caption{Mean number of wildtype mosquitoes per household over time (weeks) with a single male-only release of Wolbachia-infected mosquitoes at $150$ days. The coloured curves denote community wide releases of $800$ male mosquitoes for low (blue), moderate (red) and high (green) release dispersal rates $\gamma$. The black dashed curve represents a household release strategy where $8$ male Wolbachia-infected mosquitoes are released in each household. All other parameter values can be found in Table \ref{tab:params}, with the intermediate values for the non-release dispersal parameters.}
    \label{fig:av-wilds-hhold}
\end{figure}

Figure \ref{fig:av-wilds-hhold} shows the mean number of wildtype mosquitoes per household over time. Households reached a quasi-steady state of around $14$ wildtype mosquitoes around $7$ weeks after the simulation was initiated. The infected males were introduced after $150$ days. The wildtype population was immediately suppressed, but only temporarily. A household scale release was most effective, decreasing the mean number of wildtype mosquitoes in a household to just below $11$. A community scale release with high dispersal produced a similar result. Community scale releases with moderate and low dispersal brought the mean wildtype household population down to around $12$ mosquitoes. Under all the release scenarios, the wildtype mosquito populations recovered to the original quasi-steady state within around $15$ weeks of the release. We further observed that, on average, male Wolbachia-infected mosquitoes became extinct in the household less than $2$ weeks after the release (result is not shown here).

Clearly, regular releases of male Wolbachia-infected mosquitoes are needed in order for this strategy to be effective. A single male-only release strategy is much less effective than a single mixed sex release strategy as a consequence of the CI effect. When a Wolbachia-infected male enters a wildtype household it is unable to produce viable (and infected) offspring with the wildtype females. Consequently, infected males can dilute wildtype reproduction in the household, but cannot drive invasion of Wolbachia infection.

\subsection{The impact of mosquito sex ratios in household scale releases} \label{subsec:sex-ratios}
\begin{figure}[htbp]
    \centering
    \begin{subfigure}[b]{0.45\linewidth}
        \centering
        \includegraphics[width=\linewidth]{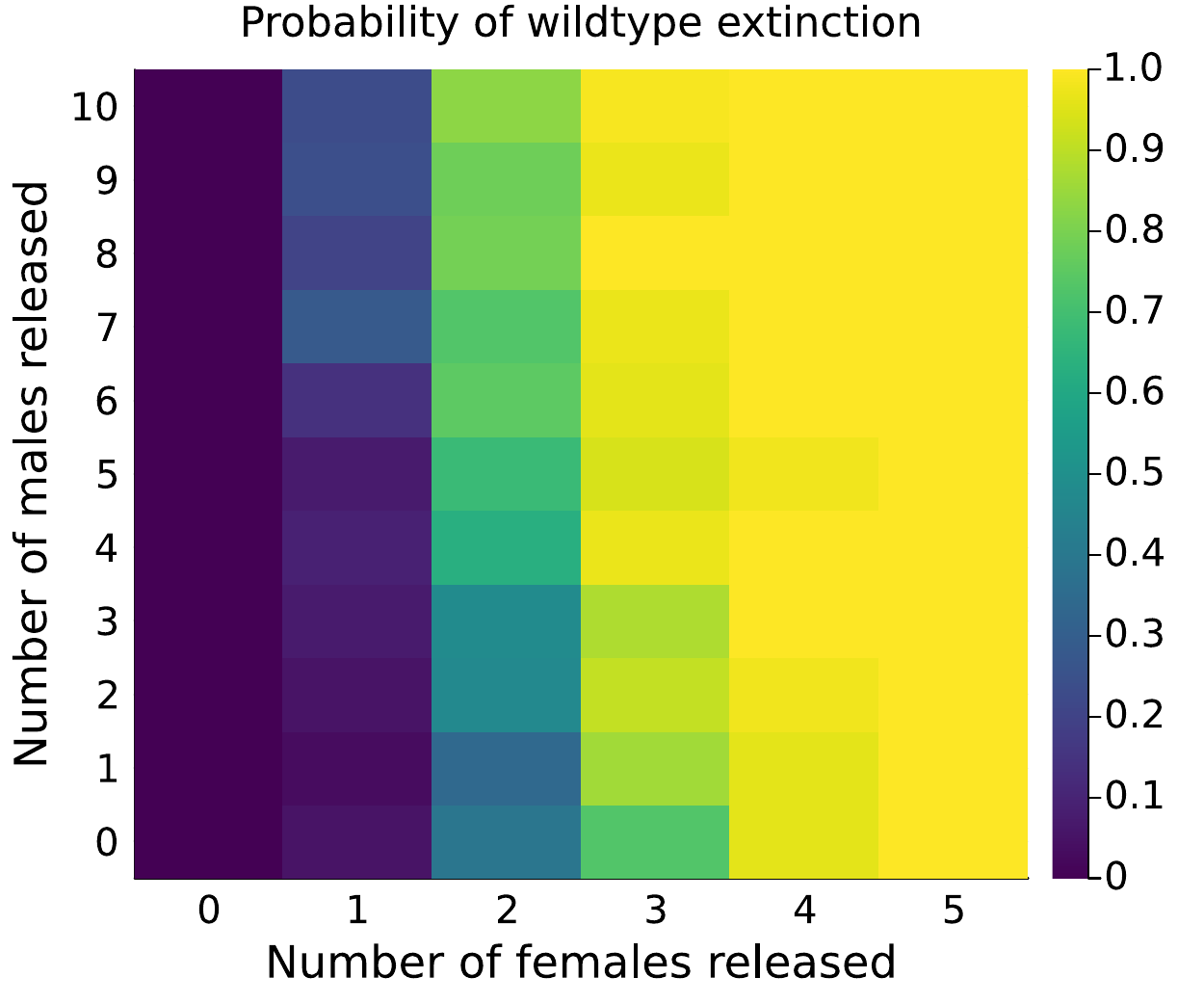}
        \caption{}
        \label{fig:inv_prob_50}
    \end{subfigure}
    \hfill
    \begin{subfigure}[b]{0.45\linewidth}
        \centering
        \includegraphics[width=\linewidth]{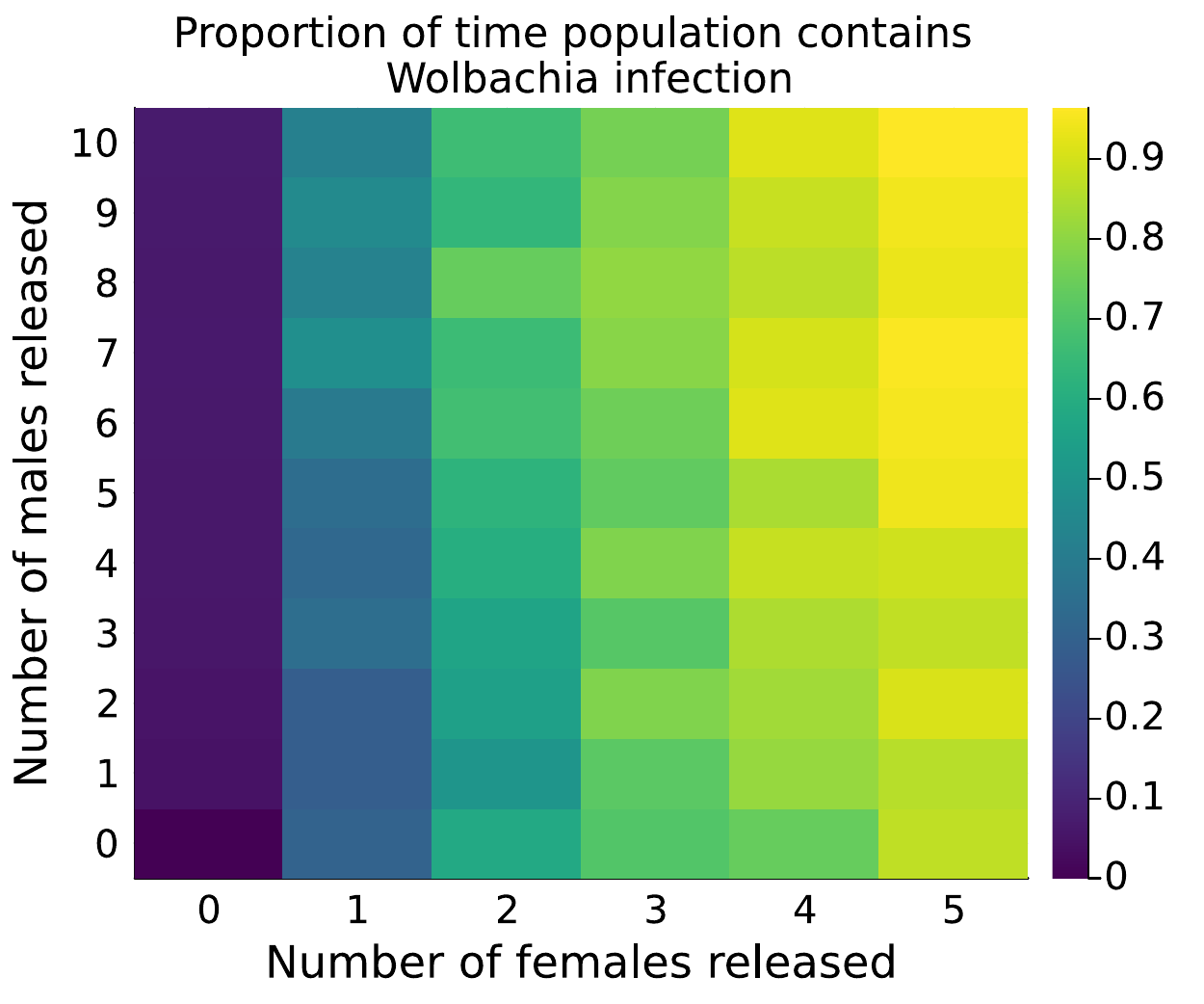}
        \caption{}
        \label{fig:pers_time_50}
    \end{subfigure}
    \caption{Outcomes after $1000$ days following a single release of $f$ female and $m$ male Wolbachia-infected mosquitoes into $50\%$ of households at $150$ days. All households were initialised with $8$ male and $8$ female wildtype mosquitoes. All parameters can be found in Table \ref{tab:params}, where we take intermediate values for the dispersal parameters. (a) Probability of wildtype extinction in the entire metapopulation. (b) Proportion of time, after day $150$, for which Wolbachia infection persists in the metapopulation.}
    \label{fig:single-mix-half}
\end{figure}

Figure \ref{fig:single-mix-half} shows the results of a single household scale release of $f$ female and $m$ male wildtypes for $m,f=0,1,2,\ldots, 10$ in half ($50$) of the households. Figure \ref{fig:inv_prob_50} shows that wildtype extinction of the whole population of households is only possible if at least one female Wolbachia-infected mosquito is released in each household. As discussed in Subsection \ref{subsec:male}, this is because under the CI effect, an infected male is unable to produce Wolbachia-infected offspring with a wildtype female. Releasing $3$ Wolbachia-infected females in each household is sufficient for metapopulation-wide Wolbachia invasion to occur with a high probability. The number of infected males released into the household only has a noticeable impact when the release includes less than $4$ infected females per household. In this case, higher numbers of Wolbachia-infected males increase the probability of wildtype extinction because infected females are more likely to produce infected offspring before dying or switching households. Lowering the rate at which female mosquitoes disperse does not qualitatively impact this result.

As indicated in Figure \ref{fig:pers_time_50}, the persistence of Wolbachia-infection improves with the number of female Wolbachia-infected mosquitoes released. When at least $5$ females are released into each household, Wolbachia-infection persists in the metapopulation for the majority of the simulation. The persistence time is insensitive to the number of males released. 

\subsection{Regular releases into a single household}\label{subsec:single-hhold}
\begin{figure}[htbp]
    \centering
    \begin{subfigure}[b]{0.45\linewidth}
        \centering
        \includegraphics[width=\linewidth]{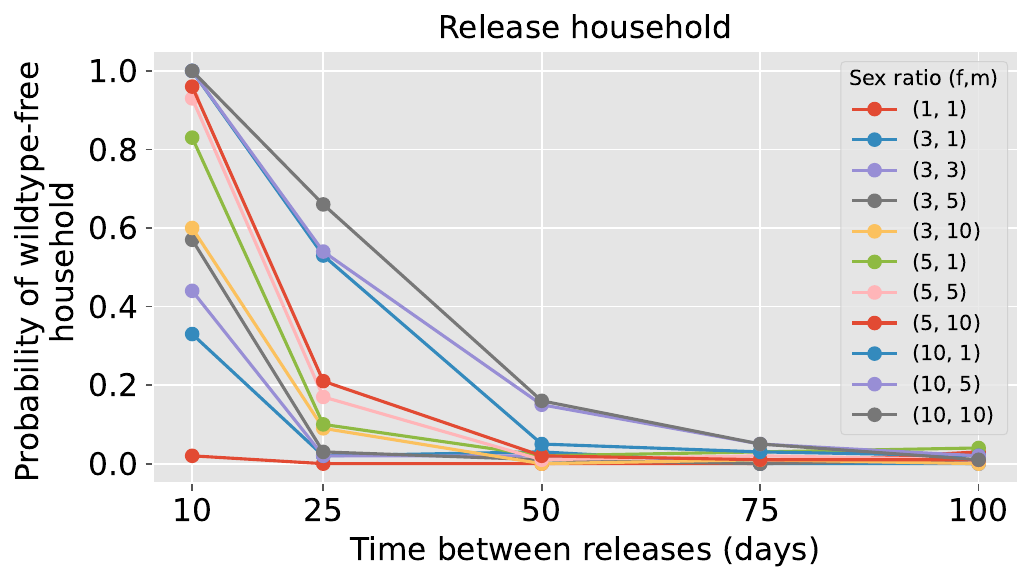}
        \caption{}
        \label{fig:inv_single}
    \end{subfigure}
    \hfill
    \begin{subfigure}[b]{0.45\linewidth}
        \centering
        \includegraphics[width=\linewidth]{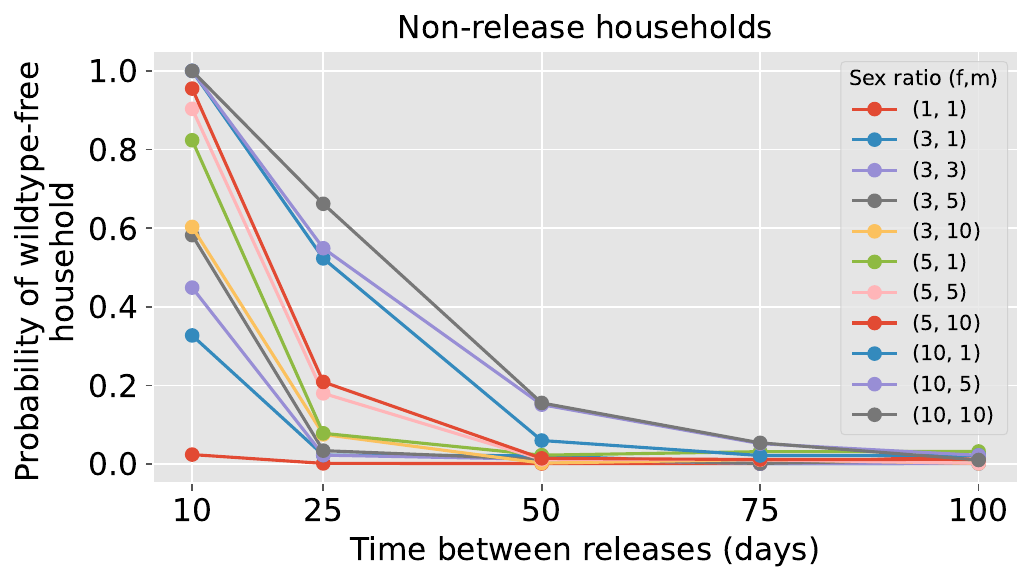}
        \caption{}
        \label{fig:inv_all}
    \end{subfigure}
    
    \vskip\baselineskip
    
    \begin{subfigure}[b]{0.45\linewidth}
        \centering
        \includegraphics[width=\linewidth]{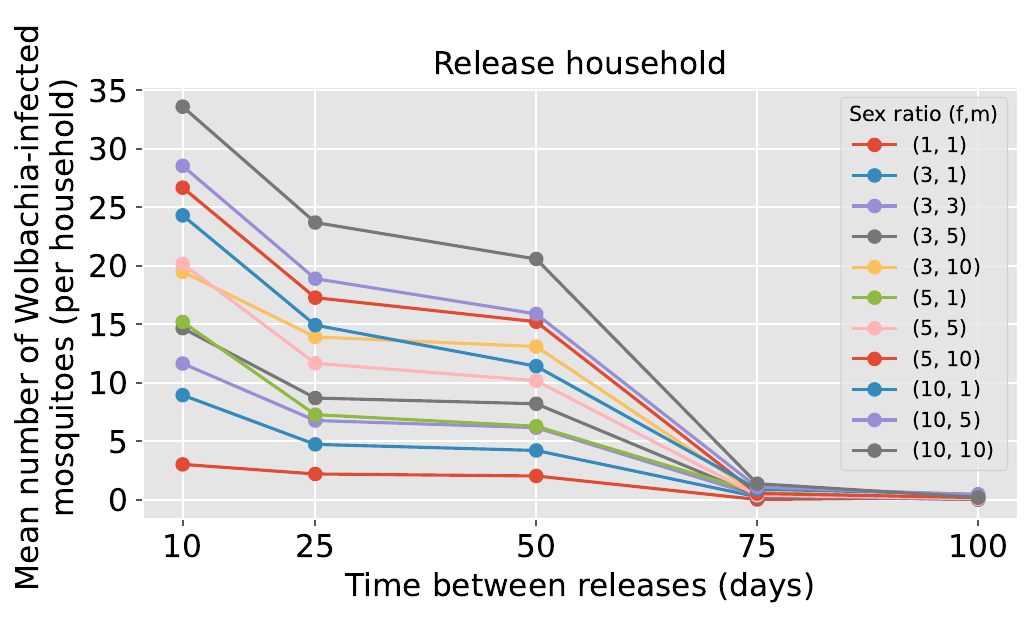}
        \caption{}
        \label{fig:wolb_size_single}
    \end{subfigure}
    \hfill
    \begin{subfigure}[b]{0.45\linewidth}
        \centering
        \includegraphics[width=\linewidth]{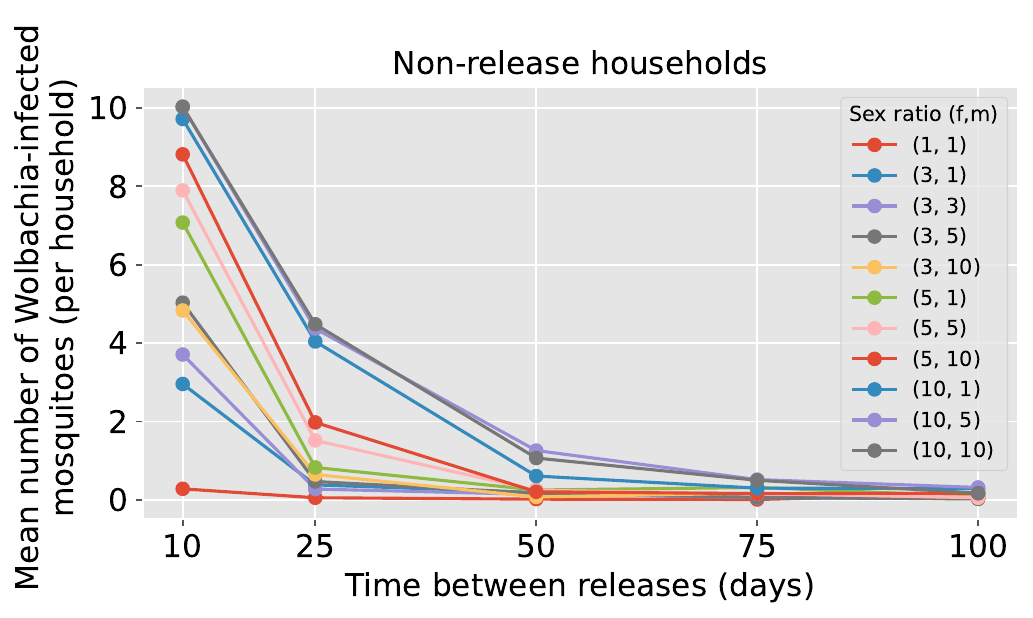}
        \caption{}
        \label{fig:wolb_size_all}
    \end{subfigure}
    
    \vskip\baselineskip
    
    \begin{subfigure}[b]{0.45\linewidth}
        \centering
        \includegraphics[width=\linewidth]{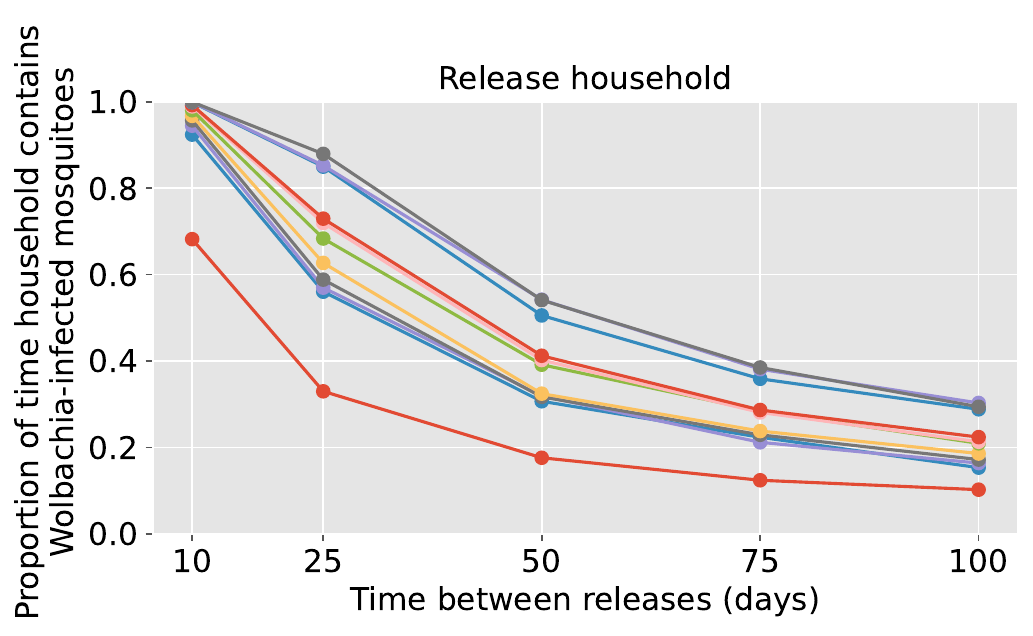}
        \caption{}
        \label{fig:pers_time_1rel}
    \end{subfigure}
    \hfill
    \begin{subfigure}[b]{0.45\linewidth}
        \centering
        \includegraphics[width=\linewidth]{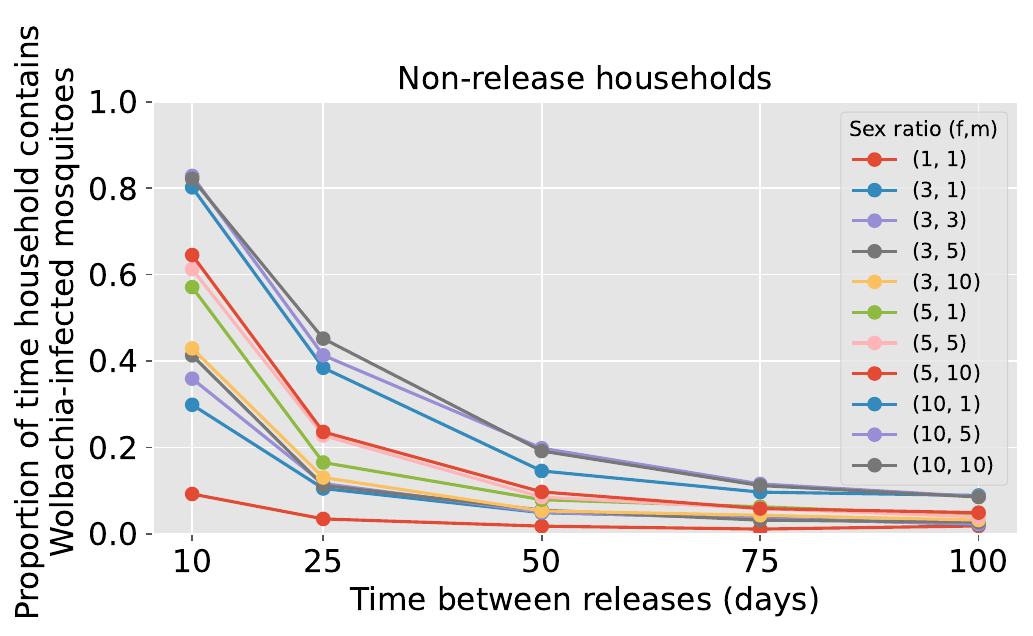}
        \caption{}
        \label{fig:pers_time_1norel}
    \end{subfigure}
    
    \caption{Outcomes of regular releases of Wolbachia-infected mosquitoes in a single household. All households were initialised with $8$ male and $8$ female wildtype mosquitoes. Starting at day $150$, Wolbachia-infected mosquitoes were introduced into the same single household with the frequency shown on the horizontal axis. Line colours indicate the numbers of male $m$ and female $f$ infected mosquitoes released $(f,m)$. (a), (b) Probability a household is wildtype-free after $1000$ days; (c), (d) Mean number of Wolbachia-infected mosquitoes in a household after $1000$ days; (e), (f) Proportion of time after day $150$ for which Wolbachia-infection is present in a household. Plots (a), (c), and (e) correspond to the household where releases occur; plots (b), (d), and (f) correspond to the other households in the community. For plots (e) and (f), the proportion of time Wolbachia-infection is present in the household is the mean derived from monitoring a single release household and a single non-release household in each of $100$ model realisations. For all other plots we average over a total of $10,000$ households ($100$ households for each $100$ model realisations). All parameters can be found in Table \ref{tab:params}, where we take intermediate values for the dispersal parameters.}
    \label{fig:single_household_2x3}
\end{figure}

Figure \ref{fig:single_household_2x3} shows results from simulations in which Wolbachia-infected mosquitoes are regularly released in a single household. Figures \ref{fig:inv_single} and \ref{fig:inv_all} show the proportion of households ($10,000$ households composed of $100$ households in each of $100$ realisations) for which the release and non-release households are wildtype-free at time $1000$ days. If Wolbachia invasion is successful in the release household then it almost always successfully invades the non-release households as well. Invasion with a probability greater than $0.8$ requires at least $5$ female Wolbachia-infected mosquitoes to be released at least every $10$ days. If $10$ infected females are released, an invasion probability greater than $0.5$ requires a release every $25$ days or less. For larger time intervals between releases or smaller release populations, the probability of a successful Wolbachia invasion is small. We hypothesise that smaller and less frequent releases will be sufficient if the proportion of households participating in releases is higher. 

Our results show that regular releases of infected mosquitoes in a single household can sustain Wolbachia-infection in an entire community. Closer inspection of the results reveals that, although the Wolbachia-infected mosquitoes spread to all non-releasing households, wildtype replacement in these households can take several years. In the release household wildtype replacement occurs within a few weeks. Wolbachia-infected mosquitoes suffer a fitness cost in terms of reduced reproduction. The CI effect can compensate for this cost, but only if the infected proportion of the household mosquito population is sufficiently high. In the release household, repeated introduction of significant numbers of infected mosquitoes generate sufficient momentum for invasion to occur with high probability. In the non-release households, invasions are typically driven by very small numbers of infected mosquitoes, and the weak momentum means that multiple invasion attempts may be required before successful establishment. When fewer female (and male) mosquitoes are released, or releases occur less frequently, it takes much longer to build sufficient momentum for successful invasion, and it may take several years for the household to reach a quasi-steady state.

In Figures \ref{fig:pers_time_1rel} and \ref{fig:pers_time_1norel} we compare the proportion of time for which Wolbachia-infection (i.e. at least one infected mosquito) is present in the release and non-release households. More frequent releases support longer persistence. In all cases, Wolbachia-infection persists for considerably longer in the release households than in households in the wider community, indicating the potential for a localised protective effect. 

Smaller time intervals between releases and larger releases of infected females produce larger numbers of Wolbachia-infected mosquitoes in the release household at time $1000$ days (see Figure \ref{fig:wolb_size_single}). When releases are carried out every $10$ days and at least $10$ females are released each time, the non-release households reach a steady state of $10$ wildtype mosquitoes, whereas the continual releases push the Wolbachia-infected population in the release household above the steady state. When releases are carried out more than $25$ days apart, the mean number of Wolbachia-infected mosquitoes in a non-release household becomes insignificant, indicating that infection is unlikely to spread from the release household (see Figure \ref{fig:wolb_size_all}). 

\section{Discussion}
In this study, we developed a model for the release and establishment of Wolbachia-infected mosquitoes at a household scale. We were motivated by the observation that Ae aegypti mosquitoes often live in urban settings, in and around the dwellings of the people that they bite \citep{dzul2017indoor}. We utilised stochastic simulation \citep{gillespie1977exact} to keep track of a dynamic set of mosquito household populations, in addition to free-living mosquito populations arising from community scale releases and mosquitoes migrating between household habitats. Our primary aim was to investigate the replacement of resident wildtype populations by introduced Wolbachia-infected populations occurring stochastically at small population scales.

We found that for mixed sex releases, releasing Wolbachia-infected mosquito populations at a household scale generally leads to more successful invasion and a higher quasi-steady state Wolbachia-infected population in the household than performing a community wide release of the same total number of mosquitoes. 

Highly localised releases are effective because the Wolbachia-infected mosquitoes suffer a fitness cost due to infection but a frequency-dependent fitness benefit due to cytoplasmic incompatibility --- the offspring of infected males and wildtype females are not viable.  Therefore, the probability of Wolbachia-infected mosquitoes successfully invading a household is higher when the initial infected population is larger. Household scale release provides localised momentum for invasion. In general, community scale releases reduce the invasion momentum because the mosquitoes disperse gradually and enter households sporadically. Community scale releases with very high dispersal rates can produce invasion dynamics comparable to household scale releases. However, we believe such a dispersal rate is improbable in urban environments. 

Community scale and household scale release strategies converge as the rate of initial dispersal into the households ($\gamma$) becomes large. We performed a sensitivity analysis for the single mixed sex release scenario, varying the dispersal rates between low, moderate and high values in all possible combinations of our four dispersal parameters. Our full set of dispersal parameters are (1) the rate at which released Wolbachia-infected mosquitoes disperse into the households ($\gamma$); (2) the rate at which male mosquitoes arrive at households from the free-living population ($\rho$); (3) the rate at which male mosquitoes leave households for the free-living population ($\tau$) and (4) the rate at which female mosquitoes switch between households ($\alpha$). The simulation data produced from our sensitivity analysis can be found in the GitHub repository \url{https://github.com/ahb48/Wolbachia-movement-households}. We observed no qualitative changes in the overall pattern of results for any of the dispersal rate parameter sets we considered. 

We also considered a male-only release of Wolbachia-infected mosquitoes with the aim of suppressing the wildtype population in the household. Although we concluded that the household scale release was once again most effective at pushing the wildtype population in the household below the steady state value, under all release scenarios the wildtype population quickly rebounded to its original steady state value following a single release event. This result suggests that if long term suppression is required, numerous releases may need to be carried out in quick succession until the wildtype population has become extinct both in the household and in the surrounding community. Alternatively, releases may be carried out on a regular basis in order to maintain the wildtype population at a lower level that reduces or eliminates the risk of vector-borne disease circulation in the community. 

The goal of the control strategy differs between mixed-sex and male-only releases and on the strain of Wolbachia used. Under the mixed-sex release, in general, the aim is to replace the wildtype population with a Wolbachia-infected population. This is because mosquitoes infected with certain strains of Wolbachia are less competent at passing on dengue infection to humans. However, the female mosquitoes will still take blood meals from the inhabitants of the household and some transmission is still possible. Under a male-only release, the goal is to suppress the total mosquito population via the infected males diluting the reproduction of mosquitoes through the CI effect. This leads to fewer (potentially no) mosquitoes biting human inhabitants and consequently lower transmission risk. It may be possible to achieve something in between these two strategies by varying the sex ratio of the Wolbachia-infected mosquitoes released. This could yield new strategies that minimise resource expenditure and disease risk. 

We explored this premise and found that releasing very small numbers of infected female mosquitoes in $50\%$ of households leads to widespread invasion of the metapopulation with high probability. The number of number of infected male mosquitoes released per household has little impact on the invasion dynamics. Therefore, in general, mixed-sex releases offer no advantage over female-only releases. 

A single release event is unlikely to be sufficient for long term control of the wildtype mosquito population, particularly if CI is imperfect or wildtype mosquitoes re-invade from regional populations. So we adapted our simulation framework to account for household scale releases occurring at regular intervals. We determined that even a single household performing releases of at least $10$ female mosquitoes at least every $10$ days, can eventually protect a whole community. However, it can take several years to achieve wildtype replacement in non-release households in this way whereas wildtype replacement in the release household is rapid. When releases are carried out less frequently or with fewer female mosquitoes, it seems possible that some bistable quasi-steady state may be reached in the household metapopulation. Similar effects, where some subpopulations in the metapopulation reach a certain steady state solution while other subpopulations find a different steady state, have been observed in deterministic ecological models \citep{vilk2020extinction}. We were not able to identify any model realisations in which some households reached a wildtype-only quasi-steady state while others reached a Wolbachia-only quasi-steady state. 

In a previous study \citep{barlow2025analysis}, we used analytical methods from probability theory to investigate the outcome of Wolbachia-infected mosquitoes invading a single household containing a resident wildtype population. Here we have extended the model to a community of households connected by the movement of mosquitoes. This dispersal introduced nonlinearities into the model that made it necessary to move towards numerical approaches. For single households, we found good agreement between our analytic and numerical frameworks. There is further discussion on this in the Supplementary Information. There is, however, a small discrepancy between the wildtype quasi-steady state number of mosquitoes in our current model and the corresponding single household steady state from \cite{barlow2025analysis}. We believe this is a consequence of the mosquito birth rate being influenced by the sex ratio when males are modelled explicitly. The dispersal of male mosquitoes further alters the household steady state since at any time a number of male mosquitoes are outside of the household, in the free-living population. We do not think this discrepancy is likely to have any qualitative impact on the model dynamics.

The numerical framework we have implemented has the additional advantage of accommodating a wide range of variation in community structure and release strategies. This flexibility can be exploited to further inform Wolbachia-based control strategies and field trials in dengue prevalent areas. In future work we aim to explore a wider range of release strategies. For instance, adaptive strategies could monitor the mosquito population sizes in the community and the household and regularly adjust the numbers of male and female mosquitoes released. Alternatively, it could be desirable to prioritise highly infested households or households with clinically vulnerable residents. These studies could reveal more efficient and cost effective control strategies. 

We acknowledge several limitations and simplifying assumptions in our framework. The size of the mosquito population in a household is restricted. Free-living mosquitoes can only enter a household that contains less than $2K$ mosquitoes in total. Mosquitoes are only born into a household if there are less than $K$ females residing there. We place a restriction on household size in order to keep the number of possible household states computationally manageable. Our bound is well within the range suggested by empirical studies \citep{madewell2019associations}. The birth rate restriction originates from our framework in \cite{barlow2025analysis} where we model the female population only. This bound is retained here verbatim for consistency. A consequence of this is that it is possible for the household to move above the household reproductive carrying capacity, $2K$, when there are less than $K$ females but $2K$ or more total mosquitoes in the household. This situation rarely occurs and we do not believe it has a significant effect on the results. 

We introduce a rescue effect into the metapopulation by allowing female mosquitoes to switch at a low rate between households, whereas male mosquitoes move from a household to a `free-living' population, and then into another household. Our framework is set up in this way to capture empirical observations of females mosquitoes dispersing over much smaller distances than males. As more data becomes available, it will be beneficial to explore the contrasting dispersal behaviours of male and female mosquitoes in more detail.

We assume vertical transmission and the CI effect are perfect, so that Wolbachia-infected females produce infected offspring with probability $v=1$ and $100$\% of offspring produced by infected males and wildtype females are unviable. Empirical studies \citep{hoffmann2011successful, xi2005wolbachia, mcmeniman2009stable} suggest this is a reasonable assumption. In our previous study \citep{barlow2025analysis}, we explored the impact of imperfect vertical transmission in Wolbachia-infected female mosquitoes. We found it did not significantly affect the dynamics of probability of Wolbachia invasion as long as $0.9\leq v < 1$.

We defined a larval density dependence function acting on the birth rate, which
describes the effect of intraspecific competition for resources such as food and space. The function is slightly adjusted in comparison to the larval density function used in \cite{barlow2025analysis} in order to account for unviable births produced via the CI effect which will not contribute towards the competition. The function is in line with that used in \cite{hughes2013modelling,dye1984models} and many other modelling studies. The model is heuristic, however data was originally used to estimate the model parameters in \citep{dye1984models} under the original scaling. Were more data to become available, it could be beneficial to revisit the model.

In conclusion, Wolbachia-infected mosquito release is a promising approach to mosquito control and consequently management of vector-borne infections such as dengue. Aedes aegypti mosquitoes are known to reside in urban areas, in and around the same dwellings as the people that they bite. In this study we accounted for stochastic effects in the invasion and establishment of Wolbachia-infected mosquitoes at a household scale and found that household scale releases are a promising avenue for efficient mosquito control.

\section{Supplementary information}
All code used to produce the results in this study can be found at \url{https://github.com/ahb48/Wolbachia-movement-households}. Supplementary Information can be found in a supporting document online.

\section{Acknowledgements}
This work was supported by a scholarship from the EPSRC Centre for Doctoral Training in Statistical Applied Mathematics at Bath (SAMBa), under the project EP/S022945/1.

For the purpose of open access, the author has applied a Creative Commons Attribution (CC-BY) license to any Author Accepted Manuscript version arising.
No new data was created during the study.




\bibliographystyle{elsarticle-harv} 
\bibliography{wolb-bibliography.bib}

\end{document}